\documentclass[aps,prb,twocolumn,showpacs,groupedaddress]{revtex4-1}
\usepackage{latexsym}
\usepackage[english]{babel}
\usepackage{graphicx}
\usepackage{amsfonts}
\usepackage{amsmath}
\usepackage{amssymb}
\usepackage{slashed}
\usepackage{verbatim}
\usepackage{feynmf}
\usepackage{mathrsfs}
\usepackage{amsmath,amssymb}
\usepackage{epsfig}
\usepackage{relsize}
\usepackage{graphicx}
\input epsf
\usepackage{bbm}
\usepackage{color}
\usepackage{bbold}
\usepackage{subcaption}
\usepackage{appendix}

\begin{document}

\title{Screening in multilayer graphene}

\author{Ralph van Gelderen}
\email{Ralphvangelderen@gmail.com}
\author{Richard Olsen}
\author{C. Morais Smith}
\affiliation{Institute for Theoretical Physics, Utrecht University, Leuvenlaan 4, 3584 CE Utrecht, The Netherlands}

\date{\today}


\begin{abstract}
In this article we study the static polarization in ABC-stacked multilayer graphene. Since the density of states diverges for these systems if the number of layers exceeds three, screening effects are expected to be important. In the random phase approximation, screening can be included through the polarization. We derive an analytical integral expression for the polarization in both the full-band model and an effective two-band model. Numerical evaluation of these integrals are very time consuming in the full-band model. Hence, for ABC-stacked trilayer graphene, we use the two-band model to calculate the low momentum part of the polarization. The results for the two-band model are universal, i.e. independent of doping. The high momentum part is linear and is determined by calculating two points, such that we can determine the slope. For ABC stacked trilayer graphene, the slope is given by three times the monolayer value. We compare our results to previous ones in the literature and discuss the similarities and discrepancies. Our results can be used to include screening in ABC-stacked multilayer systems in a way that all the characteristics of the polarization function are included. The numerical results for the polarization of trilayer graphene are used to sketch the screened potential.
\end{abstract}

\maketitle

\section{Introduction}

Stacking several layers of graphene on top of each other does not immediately lead to graphite. As long as the number of layers is small enough, the two-dimensional nature of the system is preserved, i.e. the (quasi) momentum of the particles is oriented within the plane. The properties of these systems depend heavily on the way the layers are stacked and typically differ considerably from both monolayer graphene and graphite.

\begin{figure}[t]
\includegraphics[width=.35\textwidth]{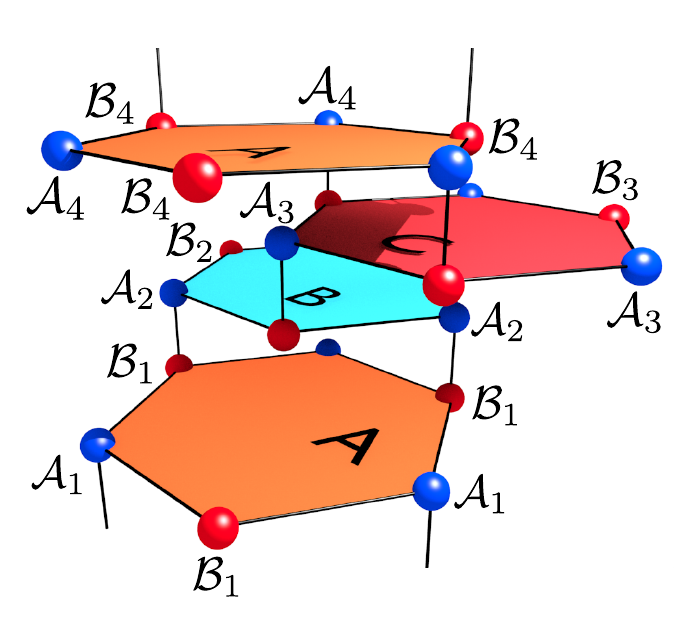}
\caption{(color online) Atomic structure of ABC-stacked multilayer graphene.}
\label{fig7.1}
\end{figure}

There are two natural ways to stack graphene layers, namely AB or Bernal stacking and ABC (rhombohedral) stacking. In Bernal stacked multilayer graphene, the odd layers all have the same orientation, and so do the even layers. The orientation of the even layers is such that the $\mathcal{B}$ sublattice sites are opposite to the $\mathcal{A}$ sublattice sites of the layers directly beneath and above it. The $\mathcal{A}$ sublattice sites are located opposite to honeycomb centers. In rhombohedral stacked graphene, every layer is oriented such that the $\mathcal{B}$ sublattice is on top of the honeycomb centers of the layer beneath it and the $\mathcal{A}$ sublattice is on top of the $\mathcal{B}$ sublattice of the layer beneath it. This results in a cyclic structure through different orientations. Hence, the layers $i$ and $i+3$ are exactly on top of each other. This lattice structure is shown in Fig.~\ref{fig7.1}.

Although a recent theoretical work investigates systems in which the stacking of the layers is partly Bernal and partly rhombohedral,\cite{Mccann12} so far most of the effort has been put into understanding either completely Bernal or completely rhombohedral stacked multilayer samples. These two systems behave very differently. In Bernal stacked multilayers, there are multiple low-energy bands, i.e. quasi particles with different dispersions. When the number of layers is even ($N=2n$), the $n$ low-energy conduction bands are all parabolic (bilayer-like), but with different effective masses, while for an odd number of layers ($N=2n+1$) a linear band with the same slope as the energy band in monolayer graphene exists next to the $n$ parabolic ones.\cite{McCa10a} The valence bands are related to the conduction bands by particle hole symmetry. On the other hand, for ABC-stacked multilayers, the low-energy physics takes place on the sublattice sites on the outer layers that do not have a direct neighbor in the next layer. As a result, it is possible to construct an effective $2 \times 2$ Hamiltonian that is valid for energies $E<<t_\perp\approx 0.3$ eV.\cite{McDo08a} From this effective model, it is easy to derive that the energy bands at small momenta and low energies disperse as $E \sim k^N$, i.e. the bands become very flat when $N$ increases. At the $K$ point, where the conduction and valence bands touch, the dispersion of the bands results into a diverging density of states when $N \geq 3$. This is in sharp contrast to Bernal stacked graphene, where the density of states never diverges at the Dirac point.

The integer quantum Hall effect could be a way to identify the different stacking orders. This is because the Landau level spectrum is very different in the two systems.\cite{Kats11} For example, in the trilayer case the Landau levels disperse with the magnetic field $B$ as $E \sim B^{3/2}$ for rhombohedral stacking, while the linear and parabolic bands in Bernal stacked trilayers give rise to two sets of Landau levels. One set disperses as a graphene monolayer, $E_\textrm{ml} \sim \sqrt{B}$, while the other behaves as $E_\textrm{bl} \sim B$, just as a graphene bilayer does. Hence, the Landau levels cross as a function of the magnetic field.\cite{McCa11} This fundamental difference is true for any $N \geq 3$ multilayer system: When the stacking is Bernal the Landau levels cross, while for ABC-stacked systems they do not (as long as the high-energy bands are neglected). Due to the low mobility of most multilayer graphene samples, much higher magnetic fields are required to observe the quantization of the Hall conductance. Nevertheless, for both Bernal and rhombohedral stacked trilayer graphene the integer quantum Hall effect is observed in experiment.\cite{TayHer11,Kumar11,Zhang11} In the Bernal stacked case, hexagonal boron nitride was used as a substrate,\cite{TayHer11} increasing the mobility by a factor of $100$. This technique may be used in the future to observe the quantum Hall effect in graphene multilayers with an even higher number of layers.

Although the integer quantum Hall effect was observed promptly, it took much longer to confirm that the fractional quantum Hall effect exists in graphene. As a result, the importance of the Coulomb interaction in graphene has long been debated. Theoretical predictions of interaction effects had been made, such as a ferromagnetic phase transition in monolayer, bilayer, and later also in trilayer graphene.\cite{PeCaNe05a,NiCaNe06a,Rvg11,Rvg13} The observation of the fractional quantum Hall effect in 2009 confirmed that interactions do play a role in graphene physics. The groups of Eva Andrei and Philip Kim reported the quantum Hall plateau in suspended graphene with high mobility at filling factor $\nu=1/3$, using a two-terminal device.\cite{Andrei09a,Kim09} However, a two-terminal setup cannot provide an unambiguous proof of the existence of the phenomenon. The issue has only been definitively settled after a four-terminal device was used to observe the $\nu=1/3$ plateau in suspended graphene\cite{Kim11} and several other plateaus at fractional filling factors in graphene on hexagonal boron nitride.\cite{Kim11a} The importance of interactions in graphene was later reiterated by other experiments, for example the renormalization of the Fermi velocity due to the Coulomb interaction in monolayers.\cite{Geim11} In addition, there is evidence that the fractional quantum Hall effect also occurs in suspended bilayer and trilayer samples.\cite{Lau10}

The Coulomb interaction is always present in systems that consist of many charged particles, like electrons. The importance of electromagnetic interactions depends on the properties of the system. The ratio of the Coulomb to kinetic energy $r_s$ is a measure of the influence that the Coulomb interaction has on the system. When the kinetic energy dominates and $r_s$ is small, the system can be described as a Fermi liquid. When $r_s$ is large new phases can occur. For a two-dimensional electron gas (2DEG), $r_s=m^* e^2/(\epsilon \hbar^2 \sqrt{\pi n_\textrm{el}})$, where $m^*$ is the effective mass of the electrons, $e$ the electron charge, $\epsilon$ the dielectric constant, and $n_\textrm{el}$ the density of electrons. Hence, for low electron densities the Coulomb interaction dominates and other phases, for example a Wigner crystal, can form. The $n_\textrm{el}^{-1/2}$ dependence is the result of the Coulomb interaction $\langle V \rangle \sim 1/ \langle r \rangle \sim \sqrt{n_\textrm{el}}$ and a quadratic kinetic energy $\langle K \rangle \sim k_F^2 \sim n_\textrm{el}$. When the dispersion is not quadratic, $r_s$ will become a different function of the electron density.

In monolayer graphene, the charge carriers behave as massless relativistic particles. Therefore, the kinetic energy scales with momentum or $\sqrt{n_\textrm{el}}$, instead of momentum squared or $n_\textrm{el}$ as it was the case in the 2DEG. Hence, the parameter $r_s$ depends only on material parameters and is independent of electron doping $r_s=e^2/(\epsilon \hbar v_F)$.\cite{SaRo10} For graphene, $\epsilon$ is the average dielectric constant of the material below and above the graphene layer, i.e. $\epsilon=1$ for suspended graphene in vacuum and $\epsilon=2.5$ for graphene on a SiO$_2$ substrate. Thus, $r_s=2.2$ and $r_s=0.8$ for these two cases, respectively. Compared to a typical 2DEG, graphene is weakly interacting. However, close to the charge neutrality point the density of states vanishes. As a consequence, there are not many electrons available for screening and the Coulomb interaction is almost unscreened, thus remaining long ranged. Indeed, the Thomas-Fermi screening vector, which is the $k \to 0$ limit of the polarization, scales with the Fermi energy and therefore vanishes if the system is close to half filling.\cite{SaRo10} In the short-wavelength limit, where $k$ is large, the polarization is linear and, in this regime, the effect of screening is a renormalization of the interaction strength.

For bilayer graphene, the parameter $r_s$ scales as $r_s \sim 1/\sqrt{n_\textrm{el}}$.\cite{SaRo10} Hence, close to half filling, where $n_\textrm{el}=0$, the interaction term should dominate the kinetic term. However, not only is it very difficult to produce a charge neutral system, due to the formation of electron hole puddles,\cite{Mart08} it is also no longer true that the Thomas-Fermi vector vanishes for $n_\textrm{el}=0$. The Thomas-Fermi vector is independent of the density of electrons in bilayer graphene. Therefore, screening is more profound in bilayers than in monolayers of graphene. The polarization can be calculated analytically and from the polarization it is possible to construct the screened potential.\cite{Gamayun2011} Due to the two-dimensional nature of the system, the potential is not exponentially screened, but remains polynomial.

For ABC-stacked multilayers with three or more layers, the density of states diverges at the charge neutrality point. Since the Thomas-Fermi vector scales with the density of states, it is expected that screening is important for such systems. Although screening in multilayer graphene has recently been studied numerically,\cite{Hwang12} we use a different approach, in which the small momentum and the large momentum parts are calculated independently. The approximation of the polarization function that we obtain in this way is valid for all momenta and low electron doping levels. Since the numerical calculation of the full polarization function is a time consuming process, our results can be used to include screening in multilayer graphene in an efficient and computational-friendly way.

The aim of this paper is to determine the polarization and the screened potential in rhombohedral stacked multilayers. Firstly, two models are introduced in section \ref{model7}. In the full-band model the full $2N \times 2N$ Hamiltonian is used, while in the two-band model an effective $2 \times 2$ matrix is introduced. In section \ref{bubble7}, the polarization is calculated in the two-band model and it is shown that this approximation breaks down for large momenta. Although we analytically derive the formal integrals which have to be solved to calculate the polarization in $N$ layers of graphene, we solve the problem numerically only for ABC-stacked trilayer graphene, as an example. We also show results for the full-band model. The screened Coulomb potentials are derived for the two-band model in section \ref{pot7} and a realistic sketch of the screened potential is drawn in the full-band model. We discuss our results in section \ref{concl7} and compare them with both Ref.~\onlinecite{Gamayun2011} and Ref.~\onlinecite{Hwang12}. Our results for the bilayer agree with the exact results derived in Ref.~\onlinecite{Gamayun2011}, but differ from the numerical ones obtained by Min et al.\cite{Hwang12} For multilayer systems, the results in Ref.~\onlinecite{Hwang12} do not display a linear regime, whereas according to our model this linear regime should become visible in the parameter range considered by them.

\section{The Model}
\label{model7}
\subsection{Full-band Hamiltonian}
To describe an ABC-stacked multilayer of graphene with $N$ layers, we use a nearest-neighbor tight-binding model. Hence, the electrons can tunnel to adjacent lattice sites within the same layer (with energy $t=3$ eV) or to direct neighbors at a distance $d=3.4$ {\AA} in other layers (with energy $t_\perp=0.35$ eV). The noninteracting tight-binding Hamiltonian in real space is given by
$$H_0=-\sum_{<l_i,l_j> , \sigma}\left( t a^{\dagger}_{l_i,\sigma}b_{l_j,\sigma}+t_\perp  a^\dagger_{(l+1)_i,\sigma} b_{l_i,\sigma}+ H.c. \right). $$
If $c \in \{a,b\}$, then $c^\dagger_{l_i,\sigma}$ ($c_{l_i,\sigma}$) creates (annihilates) an electron on lattice site $i$ in layer $l \in \{1,2,...,N\}$ with spin $\sigma \in \{ \uparrow, \downarrow \}$. \\
Since the unit cell of this system contains $2N$ lattice sites, the reciprocal space representation of this Hamiltonian is a $2N \times 2N$ matrix. After expanding around the $K$ point, the low-energy Hamiltonian is cast into the form,
\begin{align}
\nonumber  H_0 &= \int d^2 \mathbf{k} \psi^\dagger(\mathbf{k}) \mathcal{H}_0 \psi(\mathbf{k}), \\
\label{FulLHam} \mathcal{H}_0 &=\left( \begin{array}{ccccc} H_{\textrm{ml}} & B & 0& 0 &\ldots  \\ B^T & H_{\textrm{ml}} & B &0 & \ldots \\ 0 & \ddots &\ddots& \ddots &  \end{array} \right),\\
\nonumber H_{\textrm{ml}}&=\hbar v_F \left( \begin{array}{cc} 0 & k e^{i \phi(\mathbf{k})} \\ k e^{-i \phi(\mathbf{k})} &0 \end{array} \right), \\
\nonumber B &= \left( \begin{array}{cc} 0 & 0 \\ t_\perp & 0 \end{array} \right), \\
\nonumber \psi^\dagger(\mathbf{k}) &= \left( a_1^\dagger(\mathbf{k}), b_1^\dagger(\mathbf{k}), a_2^\dagger(\mathbf{k}), \ldots , b_N^\dagger(\mathbf{k}) \right),
\end{align}
where $\hbar v_F=(3/2) a t$ defines the Fermi velocity in monolayer graphene. Furthermore, $k$ is the norm of the two-dimensional momentum vector, $k=|\mathbf{k}|$, and $\phi(\mathbf{k})=\arctan\left( k_y/k_x \right)$ is the angle of the momentum vector. In the following, we refer to Eq.~\ref{FulLHam} as the \textit{Noninteracting full-band Hamiltonian}.

\subsection{Two-band Hamiltonian}

In an ABC-stacked multilayer, only the $\mathcal{A}$ sublattice in the bottom layer (layer 1) and the $\mathcal{B}$ sublattice in the top layer (layer $N$) do not have direct neighbors in an adjacent layer. The electrons on sites with a neighbor in an opposite layer will dimerise and the energy bands associated with these electrons will move away from the charge neutrality point. This results in two energy bands close to the charge neutrality point, while the other energy bands split away from zero by an energy $\sim t_\perp$.
Hence, for an ABC-stacked multilayer of graphene, the low-energy physics takes place on the $\mathcal{A}_1$ and the $\mathcal{B}_N$ sites. Therefore, it is possible to construct an effective low-energy model that takes only the two energy bands into account that are closest to the Dirac point.\cite{FanZhang10}
This low-energy Hamiltonian is a $2 \times 2$ matrix and since it takes $N$ intra-plane and $N-1$ inter-plane hoppings to go from the $\mathcal{A}_1$ to the $\mathcal{B}_N$ site, it has the form,
\begin{align}
 \nonumber H^\textrm{2B}_0 &= t_\perp \left( \frac{\hbar v_F}{t_\perp}\right)^N \int d^2\mathbf{k} \psi^\dagger_{2B}(\mathbf{k}) \mathcal{H}^{\textrm{2B}}_{0}\psi_{2B}(\mathbf{k}), \\
\label{2BHam} \mathcal{H}^{\textrm{2B}}_{0} &= \left( \begin{array}{cc} 0 & k^N e^{-i N \phi(\mathbf{k})} \\ k^N e^{i N \phi(\mathbf{k})} &0 \end{array} \right), \\
\nonumber \psi^\dagger_{2B}(\mathbf{k})&= (a_1^\dagger(\mathbf{k}),b_N^\dagger(\mathbf{k}) ).
\end{align}
We will refer to the Hamiltonian in Eq.~\ref{2BHam} as the \textit{Noninteracting two-band Hamiltonian}. This Hamiltonian is valid for small momenta at which the energies are much smaller than $t_\perp$.

\section{The Polarization Bubble}
\label{bubble7}

\begin{figure}[b]
\includegraphics[width=.33\textwidth]{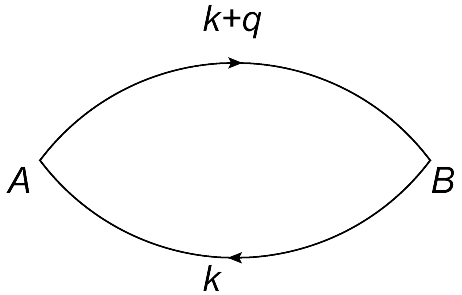}
\caption{The bubble diagram $\Pi_{AB}$.}
\label{fig7.1e}
\end{figure}

The Feynman diagram of the polarization is shown in Fig.~\ref{fig7.1e}. In the random phase approximation, this bubble diagram can be used to compute the screened potential or the free energy in an infinite order expansion. The screened part of the potential can also be absorbed into the dielectric constant.
By doing so, one can relate the polarization to the electromagnetic susceptibility $\chi(\omega,\mathbf{k})$, which is defined by $\epsilon(\omega,\mathbf{k})=1+4 \pi \chi(\omega,\mathbf{k})$, and measures the tendency of the medium to adjust to an external electromagnetic perturbation. Furthermore, the $k \to 0$ limit of the polarization gives the Thomas-Fermi screening vector. The dynamical part of the polarization is needed to describe plasmons, but those will not be treated in this paper. In one-dimensional and two-dimensional systems, the plasmon energy approaches zero and strongly couples with electrons or other quasiparticles, such as excitons. Recently, ab initio many-body calculations of the optical absorption have been performed for graphite, bilayer- and monolayer-graphene. Strong excitonic effects were found at high energy, and the results agree well with experiments in graphite.\cite{Olevano10,Yang09} In addition, exciton effects were shown to have important consequences for doped graphene systems. Indeed, theoretical studies that went beyond RPA by including electron-electron and electron-hole interactions via the many-body ab initio GW and Bethe-Salpeter equation have shown that exciton correlations enhance the cusp in the irreducible polarizability at $2 k_F$, leading to much stronger Friedel oscillations around a charged impurity than expected from RPA.\cite{Wang11} Here we neglect such excitonic effects and focus on the static polarization, for which $\omega = 0$.

\subsection{Two-band model}

In this section, we will calculate the polarization and subsequently the screened Coulomb interactions in the two-band model. We can neglect spin in this problem. The Hamiltonian is a matrix and hence, the bubble diagram will have indices labeling lattice site. For convenience, we indicate the $\mathcal{A}_1$ sites by $A$ and the $\mathcal{B}_N$ sites by $B$. The polarization is given by
 \begin{align}
 \nonumber \Pi &= \left( \begin{array}{cc} \Pi_{AA} & \Pi_{AB} \\ \Pi_{BA} & \Pi_{BB} \end{array} \right), \\
 \nonumber \Pi_{ij}(i \omega_m,\mathbf{k})&= T \sum_n \int \frac{d^2 q}{(2 \pi)^2} G_{ij}(i \Omega_n+ i \omega_m, \mathbf{q}+\mathbf{k})  \\
  &\phantom{=} \times \label{polzat}  G_{ji}(i \Omega_n,\mathbf{q}),
 \end{align}
where $G_{ij}(i \omega,\mathbf{k})$ is the electron propagator between the lattice sites $i$ and $j$, $\omega_m$ and $\Omega_n$ are Matsubara frequencies, and $T$ is temperature. In Fig.~\ref{fig7.1e} the Feynman diagram of the $\Pi_{AB}$  bubble is shown.

The propagator, which is a $2 \times 2$ matrix in the two-band model, is given by $G(i \omega,\mathbf{k})=(i \omega \mathbb{1}-\mathcal{H}_0^\textrm{2B} )^{-1}$, where $\mathbb{1}$ is the identity matrix. A derivation of the propagator in the full-band model is given in the appendix, but can be applied in the two-band model as well. The diagonalization matrices of $\mathcal{H}_0^\textrm{2B}$ are defined as
\begin{align}
\nonumber U_\mathbf{k} &= \frac{1}{\sqrt{2}} \left( \begin{array}{cc} 1 & e^{-N i \phi(\mathbf{k})} \\ e^{N i \phi(\mathbf{k})} & -1 \end{array} \right), \\
\nonumber U^\dagger_\mathbf{k} \mathcal{H}_0^\textrm{2B} U_\mathbf{k} &= D_k = \mathrm{diag}(\alpha k^N,-\alpha k^N),
\end{align}
where we have defined $\alpha \equiv t_\perp (\hbar v_F/t_\perp)^N$. Let us denote $\xi_k^s=s \alpha k^N$, where $s= \pm$. Then,
\begin{align}
\nonumber G(i \omega,\mathbf{k}) &= U^\dagger_\mathbf{k} (i \omega_m - D_k)^{-1} U_\mathbf{k}, \\
\nonumber &= \sum_s \frac{1}{i \omega_m-\xi^s_k} U^\dagger_\mathbf{k} \Delta^s U_\mathbf{k}, \\
\nonumber \Delta^+ &= \textrm{diag}(1,0), \\
\nonumber \Delta^- &= \textrm{diag}(0,1).
\end{align}
Writing $G(i \omega_m,\mathbf{k})$ in this form allows us to perform the Matsubara sum in expression (\ref{polzat}) for the polarization:

\begin{figure*}[t]
\centering
\includegraphics[width=.75\textwidth]{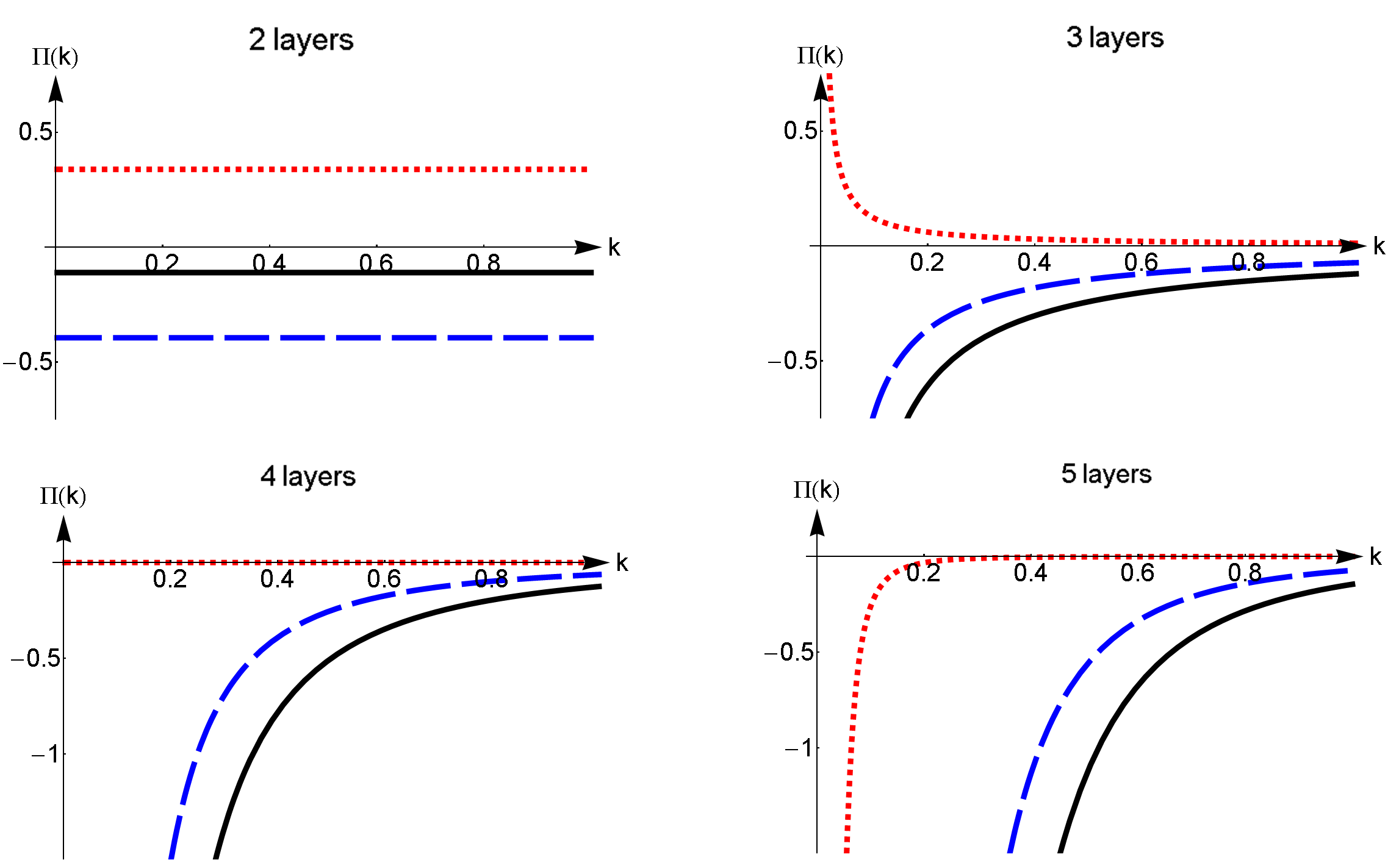}
\caption{(color online) Plot of $\alpha \Pi^0_{AA}(\omega=0,k)$ (blue dashed line), $\alpha \Pi^0_{AB}(\omega=0,k)$ (red dotted line), and $\alpha \Pi^0_{\textrm{tot}}(\omega=0,k)$ (black solid line) in the two-band model for a different number of layers and $k_F=0$.}
\label{fig7.2}
\end{figure*}

\begin{align}
\nonumber \Pi&_{ij}(i \omega_m,\mathbf{k})= \\
\nonumber &\phantom{=} T \sum_n \int \frac{d^2 q}{(2 \pi)^2} \sum_{s,s'} \frac{F_{ij}^{s s'}(k,q,\theta)}{\left(i \Omega_n+i \omega_m-\xi^s_{|\mathbf{k}+\mathbf{q}|}\right)\left( i \Omega_n-\xi^{s'}_q \right)} \\
\label{Bubbleexpr} &= \int \frac{d^2 q}{(2 \pi)^2} \sum_{s,s'} \frac{n(\xi^s_{|\mathbf{k}+\mathbf{q}|})-n(\xi^{s'}_q)}{i \omega_m+\xi_{|\mathbf{k}+\mathbf{q}|}^s-\xi^{s'}_q} F_{ij}^{s s'}(k,q,\theta), \\
\nonumber F&_{ij}^{s s'}(k,q,\theta)=\left( U^\dagger_{\mathbf{k}+\mathbf{q}} \Delta^s U_{\mathbf{k}+\mathbf{q}} \right)_{ij} \left( U^\dagger_{\mathbf{q}} \Delta^{s'} U_{\mathbf{q}} \right)_{ji}, \\
\nonumber  &= \frac{1}{4} \left\{ \delta_{ij}+s s'(1-\delta_{ij}) \cos \left[ N \phi(\mathbf{k}+\mathbf{q})-N \phi(\mathbf{q}) \right] \right\}, \\
\nonumber  &= \frac{1}{4} \left\{ \delta_{ij}+s s'\frac{(1-\delta_{ij})}{(k^2+q^2+2k q \cos \theta)^{N/2}}  \right.\\
\nonumber &\phantom{=} \times \sum_{m=0}^N \left( \begin{array}{c} N \\ m \end{array} \right) \left( q+k \cos \theta \right)^m \left( k \sin \theta \right)^{N-m}  \\
\nonumber &\phantom{=} \times \left. \cos \left( \frac{1}{2} (N-m) \pi \right) \right\},
\end{align}
where $\theta$ is the angle between $\mathbf{k}$ and $\mathbf{q}$ and $n(\xi^s)$ is the occupation function of the energy band $s$. Note that $\cos \left( \frac{1}{2} (N-m) \pi \right) \in \{-1,0,1\}$. For fixed $N$, we have determined the structure factor $ F_{ij}^{s s'}(k,q,\theta) $ analytically. In the two-band model, this factor depends on $\theta$ and on the ratio $q/k$. In the full band model we can only determine this factor numerically. Since it is (in that case) a function of $k$, $q$, and $\theta$, it will slow down the calculations.

\subsubsection*{Zero Fermi energy}

Let us first consider the half-filled system for which $E_F=0$. In this case, the valence band is completely filled, while the conduction band is empty, $$n(\xi_k^+)=0, \qquad n(\xi_k^-)=1. $$
\begin{widetext}
Let us define $x=q/k$; then the expression (\ref{Bubbleexpr}) for the polarization can be written as
\begin{align} \nonumber
  \Pi_{ij}^0(i \omega_m,k)&= \frac{-1}{k^{N-2}}\int \frac{x d x d \theta}{2 \alpha \pi^2} \frac{(1+x^2+2x \cos \theta)^{N/2}+x^N-i \omega_m/(\alpha k^N)}{\left[ (1+x^2+2x \cos \theta)^{N/2}+x^N \right]^2+\left( \frac{\omega_m}{\alpha k^N} \right)^2} F_{ij}^{s s'}(1,x ,\theta).
\end{align}
\end{widetext}
To arrive at this expression, we made the denominator real and filled in the expression for $\xi_k^s$. It is now obvious that the polarization is real, as long as $\omega_m=0$ (static screening). For static screening, one may extract the $k$ dependence of the polarization out of the integral, which will yield a constant that can be determined numerically. For the ABC-stacked trilayer, one finds that
\begin{align}
\nonumber \Pi_{AA}^0(k)=\Pi_{BB}^0(k)&=- \left(\frac{5.743}{8 \alpha \pi^2} \right)\frac{1}{k} = - \frac{0.0727}{\alpha k}, \\
\nonumber \Pi_{AB}^0(k)=\Pi_{BA}^0(k)&=\left( \frac{0.955}{8 \alpha \pi^2} \right) \frac{1}{k}= \frac{0.0121}{\alpha k}, \\
\nonumber \Pi_{\textrm{tot}}^0(k) &\equiv 2\left[ \Pi_{AA}^0(k)+\Pi_{BA}^0(k) \right]\\
\nonumber &=-\left( \frac{9.576}{8 \alpha \pi^2} \right) \frac{1}{k}  = -\frac{0.1213}{\alpha k}.
\end{align}
We conclude that for an undoped trilayer, the polarization goes as $\sim 1/k$ and therefore diverges as $k \to 0$. This is expected, because the long wavelength limit of the bubble is proportional to the density of states at the Fermi energy and for an ABC-stacked trilayer the density of states diverges at the Dirac point. This singular behavior is present for all ABC-stacked multilayers with $N \ge 3$, as can be seen in Fig.~\ref{fig7.2}.

\subsubsection*{Nonzero Fermi energy}

Now, let us assume a nonzero Fermi energy $E_F>0$. In this case, the conduction band is partially filled. Hence, the occupation functions become $$n(\xi_k^+)=\Theta(k_F-k), \qquad n(\xi_k^-)=1, $$
where $\Theta(x)$ is the Heaviside theta function and $k_F$ is the Fermi momentum. If we write down the expression for the polarization, we recognize the expression for the half filled case, $\Pi^0_{ij}(k)$ plus a correction

\begin{widetext}
\begin{align}
\nonumber  \Pi_{ij}(i \omega_m,k)&= \int \frac{q d q d \theta}{4 \alpha \pi^2} \bigg\{ \frac{\left[n(\xi^+_{|\mathbf{k}+\mathbf{q}|})-n(\xi^-_q)\right] F_{ij}^{+-}(k,q,\theta)}{(k^2+q^2+2 k q \cos \theta)^{N/2}+q^N+\frac{i \omega_m}{\alpha k^N}}- \frac{\left[n(\xi^-_{|\mathbf{k}+\mathbf{q}|})-n(\xi^+_q)\right] F_{ij}^{-+}(k,q,\theta)}{(k^2+q^2+2 k q \cos \theta)^{N/2}+q^N-\frac{i \omega_m}{\alpha k^N}} \\
\nonumber &\phantom{=}+ \frac{\left[n(\xi^+_{|\mathbf{k}+\mathbf{q}|})-n(\xi^+_q)\right] F_{ij}^{++}(k,q,\theta)}{(k^2+q^2+2 k q \cos \theta)^{N/2}-q^N+\frac{i \omega_m}{\alpha k^N}} \bigg\} \\
\nonumber &= \Pi^0(\omega_m,k)_{ij}+ \frac{1}{k^{N-2}}\int \frac{x d x d \theta}{4 \alpha \pi^2} \bigg\{ \frac{\Theta(k_F/k-\sqrt{1+x^2+2x \cos \theta}) F_{ij}^{+-}(1,x,\theta)}{(1+x^2+2x \cos \theta)^{N/2}+x^N+\frac{i \omega_m}{\alpha k^N}} \\
\nonumber &\phantom{=}+ \frac{\Theta(k_F/k-x) F_{ij}^{-+}(1,x,\theta)}{(1+x^2+2 x \cos \theta)^{N/2}+x^N-\frac{i \omega_m}{\alpha k^N}} + \frac{\left[\Theta(k_f/k-\sqrt{1+x^2+2x \cos \theta})-\Theta(k_F/k-x) \right] F_{ij}^{++}(1,x,\theta)}{(1+x^2+2 x \cos \theta)^{N/2}-x^N+\frac{i \omega_m}{\alpha k^N}} \bigg\}.
\end{align}
\end{widetext}

The term with $F_{ij}^{--}(1,x,\theta)$ vanishes, because both occupation functions are unity and therefore cancel each other out. From the expression above, we learn that only the extra terms with respect to the undoped case depend on the Fermi momentum. These correction terms contain two Heaviside theta functions. One of them, $\Theta(k_F/k-x)$, is nonzero only within a circle of radius $\delta=k_F/k$ around the origin, while the other one, $\Theta(k_F/k-\sqrt{1+x^2+2x \cos \theta})$, is nonzero within a distance $\delta$ from the point $(1,0)$. Since $\delta \to 0$ when $k \to \infty$, these correction terms will go to zero for large momenta. We conclude that, within the two band model, the large momentum dependence of the polarization always equals that of the half-filled system. Note that for $k_F/k<1/2$, i.e. $k>2k_F$, the two regions where the Heaviside functions are nonzero do not overlap. For $k<2 k_F$ these two regions do overlap and this explains the cusp in the polarization at $k=2k_F$.

\begin{figure}[b]
\centering
\begin{subfigure}[t]{0.47\textwidth}
\centering
\includegraphics[width=.95\textwidth]{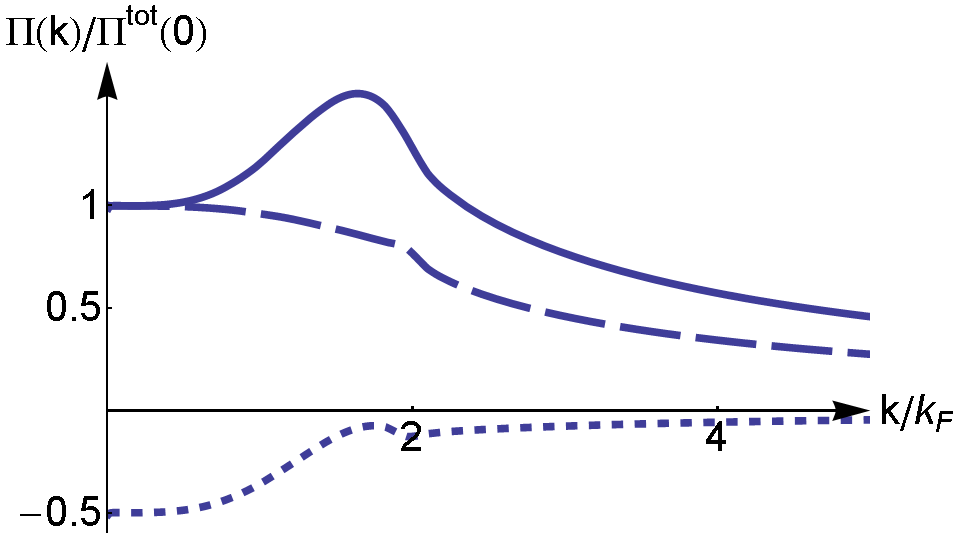}
\caption{}
\label{fig7.3a}
\end{subfigure}
\begin{subfigure}[t]{0.47\textwidth}
\centering
\includegraphics[width=.95\textwidth]{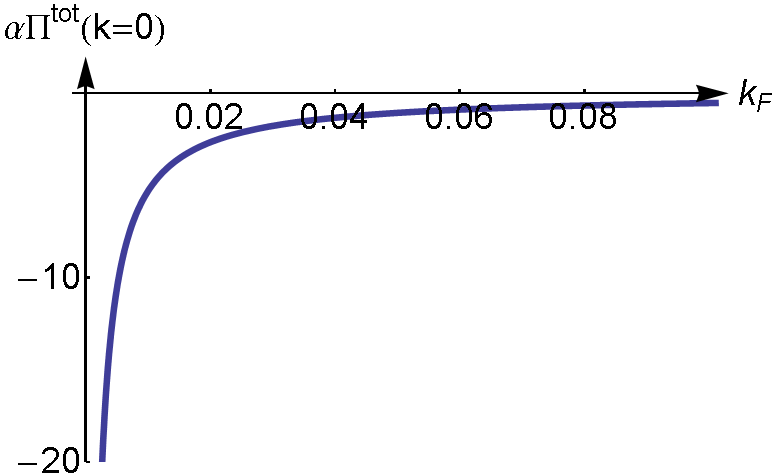}
\caption{}
\label{fig7.3b}
\end{subfigure}
\caption{(a) The normalized polarization functions for ABC-stacked trilayer graphene in the two-band model: $\Pi_{AA}(k)$ (dashed line), $\Pi_{AB}(k)$ (dotted line), and $\Pi^\textrm{tot}(k)$ (solid line). This plot is universal as long as $k_F \neq 0$. (b) Plot of $\alpha \Pi^\textrm{tot}(\omega=0,k=0)$ as a function of $k_F$.}
\label{fig7.3}
\end{figure}

The long wavelength limit of the polarizations are given by \begin{align} \label{lowenlimpol} \Pi^\textrm{tot}(k \to 0)=-\frac{g_D}{2 \pi} k_F \bigg| \frac{d \xi_p}{dp} \bigg|_{p=k_F}^{-1}, \end{align} where $g_D$ is the degeneracy factor due to spin and valley. This condition is indeed satisfied by the numerical results we have found for the trilayer case. In Fig.~\ref{fig7.3a} the components $\Pi_{ij}(k)$ and $\Pi^\textrm{tot}(k)$ are plotted for a nonzero Fermi momentum in the ABC-stacked trilayer. After scaling the axes as is done in the figure, the plot is universal, i.e. independent of Fermi momentum.

\subsection{The full-band model}

For the full-band model, the expression for the polarization can be derived in a similar way as we did to obtain Eq.~(\ref{Bubbleexpr}). The main difference is that all the matrices are now $2N \times 2N$ and in general, cannot be calculated analytically anymore. A derivation for the propagator is shown in the appendix. We have that
\begin{align}
\nonumber \Pi&_{ij}(i \omega_m,\mathbf{k})= \\
\nonumber &\phantom{=} T \sum_n \int \frac{d^2 q}{(2 \pi)^2} G_{ij}(i \Omega_n+ i \omega_m, \mathbf{q}+\mathbf{k}) G_{ji}(i \Omega_n,\mathbf{q}), \\
\nonumber &= T \sum_n \int \frac{d^2 q}{(2 \pi)^2} \sum_{s,s'} \frac{F_{ij}^{s s'}(k,q,\theta)}{\left(i \Omega_n+i \omega_m-\xi^s_{|\mathbf{k}+\mathbf{q}|}\right) \left( i \Omega_n-\xi^{s'}_q \right)} , \\
\nonumber &= \int \frac{d^2 q}{(2 \pi)^2} \sum_{s,s'} \frac{n(\xi^s_{|\mathbf{k}+\mathbf{q}|})-n(\xi^{s'}_q)}{i \omega_m+\xi_{|\mathbf{k}+\mathbf{q}|}^s-\xi^{s'}_q} F_{ij}^{s s'}(k,q,\theta), \\
\nonumber F&_{ij}^{s s'}(k,q,\theta)=\left( U^\dagger_{\mathbf{k}+\mathbf{q}} \Delta^s U_{\mathbf{k}+\mathbf{q}} \right)_{ij} \left( U^\dagger_{\mathbf{q}} \Delta^{s'} U_{\mathbf{q}} \right)_{ji},
\end{align}
where $i,j$ now label the $2N$ different sublattice sites, $s$ and $s'$ label the $2N$ energy bands, and therefore $\Delta^s$ are matrices with zero's everywhere, except for a $1$ in the $s^\textrm{th}$ entry along the diagonal. $U_\mathbf{k}$ is the diagonalization matrix of Hamiltonian~(\ref{FulLHam}). For the full-band model we define $\Pi^\textrm{tot}(k)=\sum_{i,j=1}^{2N} \Pi_{ij}(k)$.

The computation of the static polarization ($\omega=0$) is very tedious if the number of layers is greater than three. We can argue how $\Pi^{\textrm{tot}}(k)$ behaves in the short- and long-wavelength limit. The relation~(\ref{lowenlimpol}) still holds and therefore we can plot $\Pi(\omega=0,k=0)$ as a function of Fermi momentum (Fig.~\ref{fig7.3b}). As expected, in the small-$k$ limit the total polarization of the full model agrees with the results we found in the two-band model.

\begin{table}[h]
\centering
\begin{tabular}{|l|l|}
 \hline $k/k_F$ & $\Pi^\textrm{tot}(k)$\\
 \hline
 \hline 6 & 0.018  \\
 \hline 7 & 0.021  \\
 \hline
\end{tabular}
\caption{Numerical values of $\Pi^\textrm{tot}(k)$ in ABC-stacked trilayer graphene for $k_F=0.017$. }
\label{tab7.1}
\end{table}

\begin{figure}[t]
\begin{subfigure}[t]{0.47\textwidth}
\includegraphics[width=.95\textwidth]{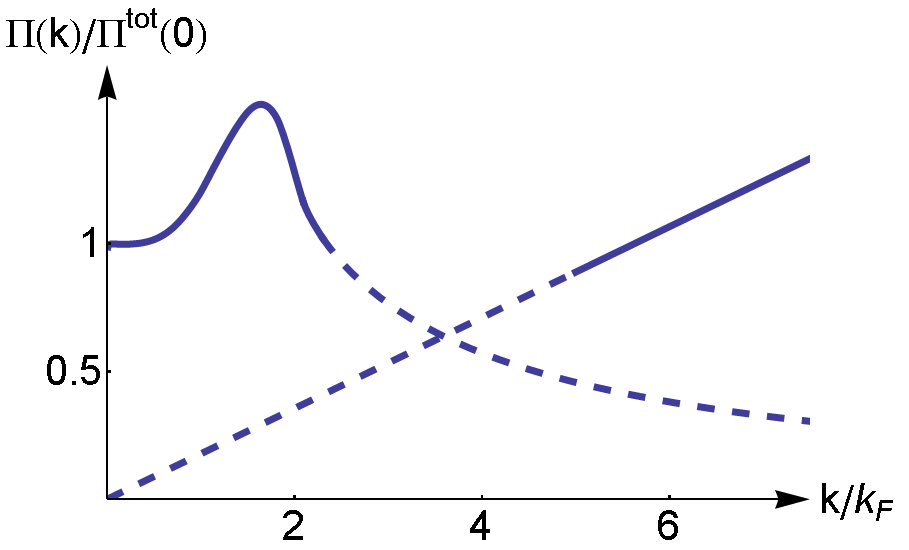}
\caption{}
\label{fig7.6a}
\end{subfigure}
\begin{subfigure}[t]{0.47\textwidth}
\includegraphics[width=.95\textwidth]{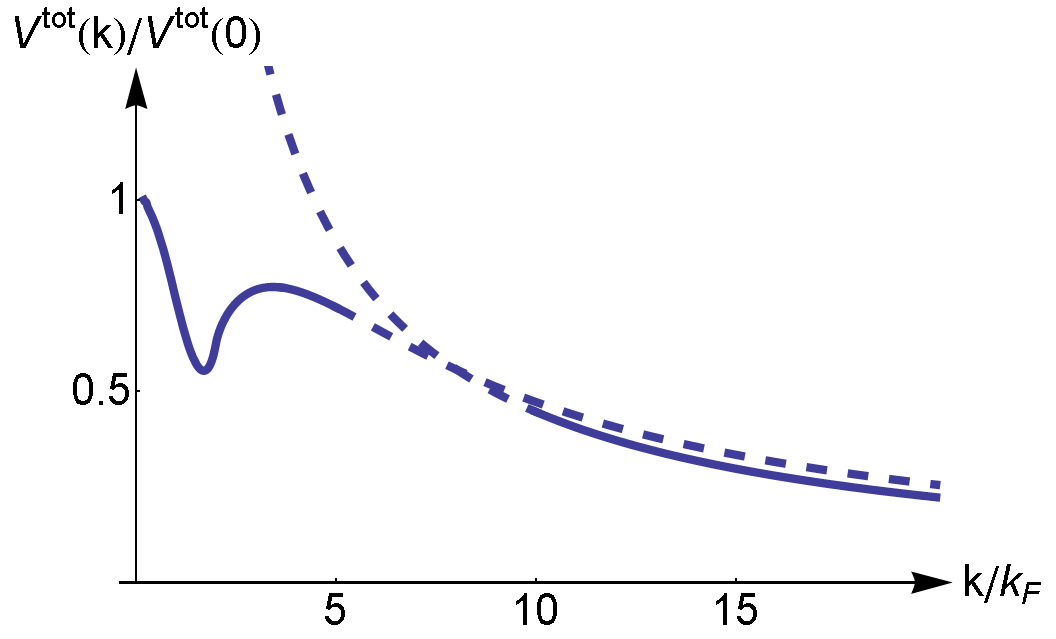}
\caption{}
\label{fig7.6b}
\end{subfigure}
\caption{The solid lines sketch (a) the full-band polarization and (b) the full-band screened potential $V^\textrm{tot}(k)$ for ABC-stacked trilayer graphene. The low-$k$ regime is obtained in the two-band model and the high-$k$ regime is obtained through a direct numerical computation. (a) The full-band polarization for $E_F/\alpha=0.027$ ($k_F=0.3$). (b) The full-band screened potential for $E_F/\alpha=0.001$ ($k_F=0.1$). In the high-$k$ regime the potential becomes a rescaled version of the unscreened potential $V(k)=1/k$.}
\label{fig7.6}
\end{figure}

Although the two band model describes the polarization well for small momenta, the short-wavelength behavior differs dramatically. We found that in the two-band model $\Pi^\textrm{tot}(k \to \infty) \sim 1/k^{N-2}$. This relation was induced by the $k^N$ dispersion. In the full band model, this dispersion relation is only valid for momenta close to the Dirac point. For larger momenta, the dispersion of the bands eventually becomes linear. If $k$ is large, this linear regime of the dispersion dominates the polarization integral and therefore $\Pi(k \to \infty) \sim k$, as is the behavior for monolayer graphene.\cite{Hwang07} Hence, the short-wavelength limit of the static bubble is linear. This linearity is independent of $N$. The slope does depend on the number of layers, but not on the Fermi energy. For trilayer graphene we have confirmed the linear behavior for a fixed Fermi energy ($k_F=0.017$) through a straight forward numerical calculation. This very time consuming process resulted in two points that align perfectly with the origin (see Table~\ref{tab7.1}), confirming that $\Pi^\textrm{tot}(k \to \infty) = -\gamma k$, with $\gamma\approx 0.18$. In Fig.~\ref{fig7.6a}, the two limiting regions of $\Pi^\textrm{tot}(k)$ are shown. The low momentum behavior, computed in the two-band model, is universal after scaling. Since the slope of the linear part is fixed, it is not invariant after scaling and therefore depends on $k_F$. Hence, the total polarization in the full-band model is no longer universal.

In this paper, we concentrated on the behavior of the polarization in the static limit ($\omega_m \to 0$). A generalization to include the dynamical part would be an interesting topic for further studies.

\section{The screened potentials}
\label{pot7}
\subsection{Two-band model}
Now that we have obtained the polarization functions, it is possible to determine the screened Coulomb potentials. In the two band model we only have
\begin{figure}[b]
\includegraphics[width=.4\textwidth]{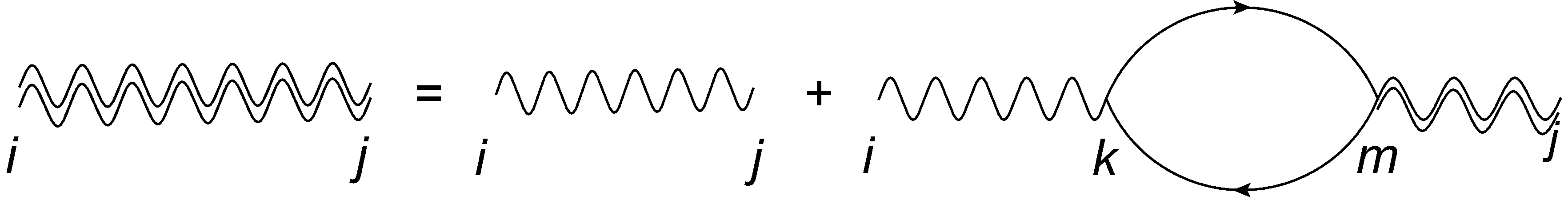}
\caption{The Dyson equation, which is a matrix equation.}
\label{fig7.4}
\end{figure}
the potentials within the layers and the potential between the first and $N$-th layer, since the low-energy physics takes place on the $\mathcal{A}_1$ and $\mathcal{B}_N$ sublattice sites. It is difficult to estimate the results for the nearest-layer potentials, $V_{\mathcal{A}_1 \mathcal{B}_2}(k)$ for example, since we cannot say anything about the nearest-neighbor polarizations, e.g. $\Pi_{\mathcal{A}_1 \mathcal{B}_2}$. In fact, in the full-band model we need all the components of the polarization matrix, if we want to calculate any screening potential. Nevertheless, let us use the two band model for now and discuss its limitations later.

\begin{figure}[t]
\centering
\includegraphics[width=.45\textwidth]{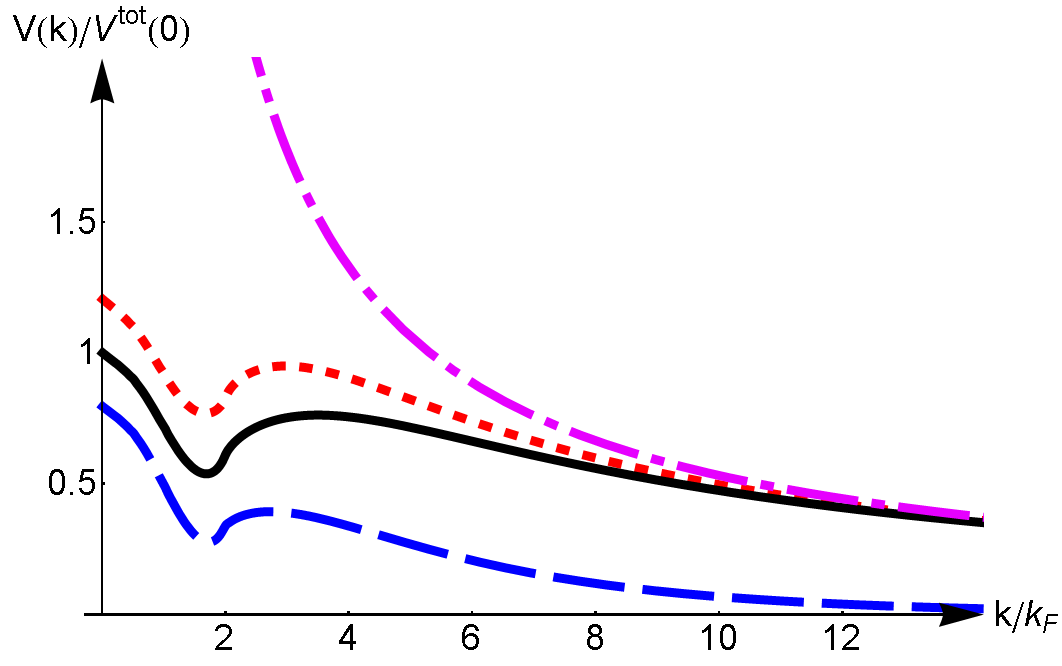}
\caption{(color online) The screened potentials in the two-band model: $V_{AA}$ (red dotted line), $V_{AB}$ (blue dashed line), and $V^\textrm{tot}$ (black solid line). The unscreened interaction $V(k)=1/(k)$ is given by the purple dotdashed line. The interlayer distance $d$ is set equal to $1$ and $k_F=0.1$.}
\label{fig7.5}
\end{figure}

The screened potentials are solutions of a Dyson equation (see Fig.~\ref{fig7.4}). However, since the potentials carry two layer indices, the Dyson equation is a matrix equation in this case. We focus now on the trilayer case. If one writes down the equations for the intralayer potential $V_{\mathcal{A}_1\mathcal{A}_1}(k)=V_{\mathcal{B}_3\mathcal{B}_3}(k)\equiv V_{AA}(k)$ and the interlayer potential $V_{\mathcal{A}_1\mathcal{B}_3}(k) \equiv V_{AB}(k)$, one notices that they are coupled. By defining $\Upsilon$ as
\begin{align}
\nonumber \Upsilon(\mathbf{k})=& \left\{ [V_{AA}(k)-V_{AB}(k)][\Pi_{AA}(k)-\Pi_{AB}(k)]-1 \right\}  \\
\nonumber & \times \left\{ [V_{AA}(k)+V_{AB}(k)][\Pi_{AA}(k)+\Pi_{AB}(k)]-1 \right\},
\end{align}
the solutions are given by
\begin{align}
\nonumber V_{11}(k)=\frac{V_{AA}(k)-\Pi_{AA}(k)[V_{AA}(k)^2-V_{AB}(k)^2]}{ \Upsilon(\mathbf{k})}, \\
\nonumber V_{13}(k)=\frac{V_{AB}(k)-\Pi_{AB}(k)[V_{AB}(k)^2-V_{AA}(k)^2]}{\Upsilon(\mathbf{k})},
\end{align}
where the bare interactions read
\begin{align*}
V_{AA}(k)&=\frac{1}{k}, \\
V_{AB}(k)&=\frac{e^{-2 k d}}{k},
\end{align*}
and $d$ is the interlayer distance.\footnote{Although we have defined the diagonalizers of the Hamiltonian, we did not make a change of basis. The (unscreened) potentials do have their usual form.} The screened potentials are shown in Fig.~\ref{fig7.5}. For convenience we have also defined $V^{\textrm{tot}}(k)=1/[k+\Pi^{\textrm{tot}}(k)]$, which is the solution to the Dyson equations if we set all exponentials $\exp(-2kd)$ equal to unity, which is a valid approximation when $k$ is small. $V^{\textrm{tot}}(k)$ has the same features as the real solutions, but because of its much simpler form, it will be convenient to use this potential during more in depth discussions.

Although this approximation describes well the screening for small momenta, the large $k$ behavior of the polarization in the full band model is different than the $1/k^{N-2}$ decay in the two-band model. Hence, the screened potentials in the two-band model will also be incorrect in the large-$k$ limit.

\subsection{Full-band model}

The screened potentials in the full-band model are also solutions of the Dyson equation sketched in Fig.~\ref{fig7.4}, which, for trilayer graphene, is now a $6 \times 6$ matrix equation. The solution is given by $$V^\textrm{scr}=\left( \mathbb{1} - V \Pi \right)^{-1} V.$$
It is easy to obtain the screened potentials numerically, once all components of the polarization are known. However, as we have seen previously, the calculation of the components of $\Pi$ is numerically very time consuming in the full-band model. Nevertheless, we obtained a realistic sketch of $\Pi^\textrm{tot}(k)=\sum_{i,j=1}^{2N} \Pi_{ij}(k)$. As before, when we put the exponentials in the potentials to unity (equivalently put $d=0$), the screened potential reduces to $V^{\textrm{tot}}(k)=1/[k+\Pi^{\textrm{tot}}(k)]$. Therefore we can sketch $V^\textrm{tot}(k)$ as well, which is done in Fig.~\ref{fig7.6b}.

\section{Discussions}
\label{concl7}

In this paper we have derived an expression for the polarization for ABC-stacked multilayer graphene. The calculations were performed within both the full-band model and the two-band model, an effective low-energy model in which the $2N \times 2N$ matrices reduce in size to $2 \times 2$. The advantage of the effective model is that it becomes easier to calculate the polarization numerically. The drawback is that the large-$k$ behavior of the polarization becomes flawed. Instead of the linear behavior $\Pi \sim -\gamma k$ at large $k$, which is imposed by the linearity of the energy bands farther away from the Dirac points, the polarization drops off as $1/k^{N-2}$ in the two-band model. It is very time consuming to calculate the full-band polarization numerically, such that the results of the two-band model are of great help to understand the behavior of $\Pi^\textrm{tot}(k)=\sum_{i,j=1}^{2N} \Pi_{ij}(k)$ in the full-band model. One has to be aware that in the effective model only the $\mathcal{A}_1$ and $\mathcal{B}_N$ lattice sites are involved. Therefore, we can say nothing about the polarization functions between other lattice sites in this approximation.

We have confirmed the linear behavior of $\Pi^\textrm{tot}(k)$ for the ABC-stacked trilayer by calculating two points in the full-band model at high momenta. These two points align with the origin, confirming the linear asymptote. Although we do not know the exact cross over between the low momentum and the linear regime, we can sketch the total polarization (see Fig.~\ref{fig7.6a}). The curves obtained in the two-band model for the polarization are universal after scaling, the cross over to the linear regime is not. This is because the slope of the linear regime is constant. After scaling however, the slope becomes $-\gamma/\Pi^\textrm{tot}(0)$. Since $|\Pi^\textrm{tot}(0)|$ is decreasing as a function of $k_F$ (see Fig.~\ref{fig7.3b}), the linear asymptote will become steeper in the scaled plots when $k_F$ is increased. For example, the cross over to the linear regime is around $2 k_F$ for $E_F/\alpha =0.03$ ($k_F=0.3$), while it is around $8 k_F$ for $E_F/\alpha =0.001$ ($k_F=0.1$). Hence, the polarization is no longer universal after scaling in the full-band model.

\begin{figure}[t]
\centering
\includegraphics[width=.45\textwidth]{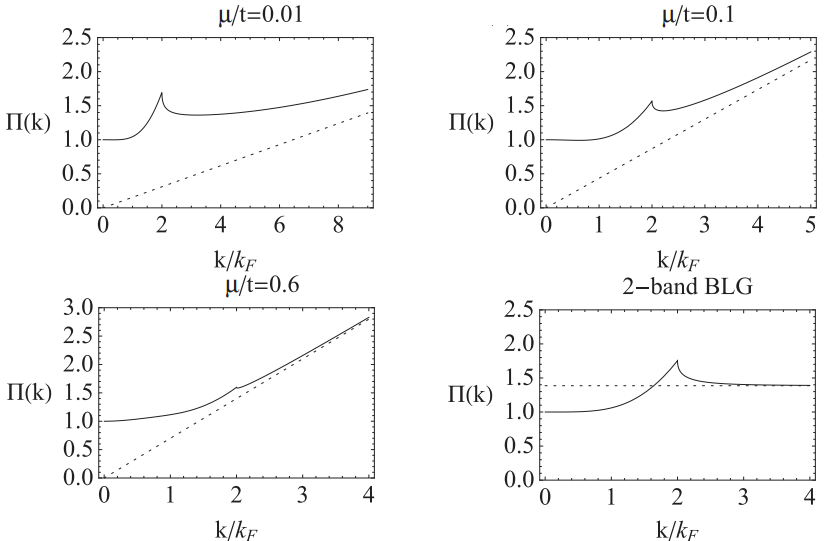}
\caption{Rescaled polarization ($\Pi^\textrm{tot}(k)$) for the graphene bilayer. Picture extracted from Ref.~\onlinecite{Gamayun2011}. $\mu$ is the Fermi energy and $t$ equals $t_\perp$ in the notation used throughout this paper.}
\label{fig7.7}
\end{figure}

Comparing our results for the ABC-stacked trilayer with the full-band\cite{Gamayun2011} and two-band\cite{Hwang08} results for the bilayer, which are shown in Fig.~\ref{fig7.7}, we see some similarities. As long as the Fermi energy is nonzero the total polarization is finite everywhere. The $k \to 0$ limit is proportional to the density of states, just as in the bilayer case. This is in fact true for any $N$-layer ABC-stacked sample. It is important to realize that, although the plots look similar, the absolute values of $\Pi^\textrm{tot}(k \to 0)$ are much higher for the ABC-stacked trilayer than for the bilayer. This is because the density of states is much higher. The normalized polarization has a peak as $k$ increases and then a cross over to a linear regime takes place if $k$ is further increased. This cross over is not captured by the two-band model. There are also differences in the behavior of bilayer and trilayer samples. The peak in the bilayer is located exactly at $2 k_F$ and has a discontinuity in the first derivative. For the trilayer the peak is smoother and the maximum is reached before $2 k_F$. The discontinuity in the first derivative at $2 k_F$ remains, although it is less pronounced. Also for the bilayer the crossing to the linear regime is shifting to smaller momenta (measured in units of $k_F$) when the Fermi momentum increases. The slope of the linear part is different for the trilayer compared with the bilayer (and the monolayer). This can be explained by the existence of more valence bands which are filled in the trilayer. In general, for an ABC-stacked $N$-layer system, there are $N-1$ filled valence bands (and also $N-1$ empty conduction bands) further away from zero energy. These bands are expected to give a contribution when computing the full-band polarization. Hence, there is no reason that the slope of the linear asymptote should be independent of layer number $N$. We expect that the slope of the linear part of the total polarization increases further when $N$ grows larger. The static polarization for monolayer graphene is constant up to $k=2k_F$ and then becomes linear with a slope $\gamma_\textrm{m.l.}=1/16$.\cite{Hwang07} Comparing this with the slope of the bilayer $\gamma_\textrm{b.l.}=1/8=2/16$, and the trilayer $\gamma=0.18 \approx 3/16$, a trend is seen. Although our results are not accurate enough, this is an indication that the slope of the linear part of the polarization scales linearly with the number of layers.

\begin{figure}[t]
\centering
\includegraphics[width=.45\textwidth]{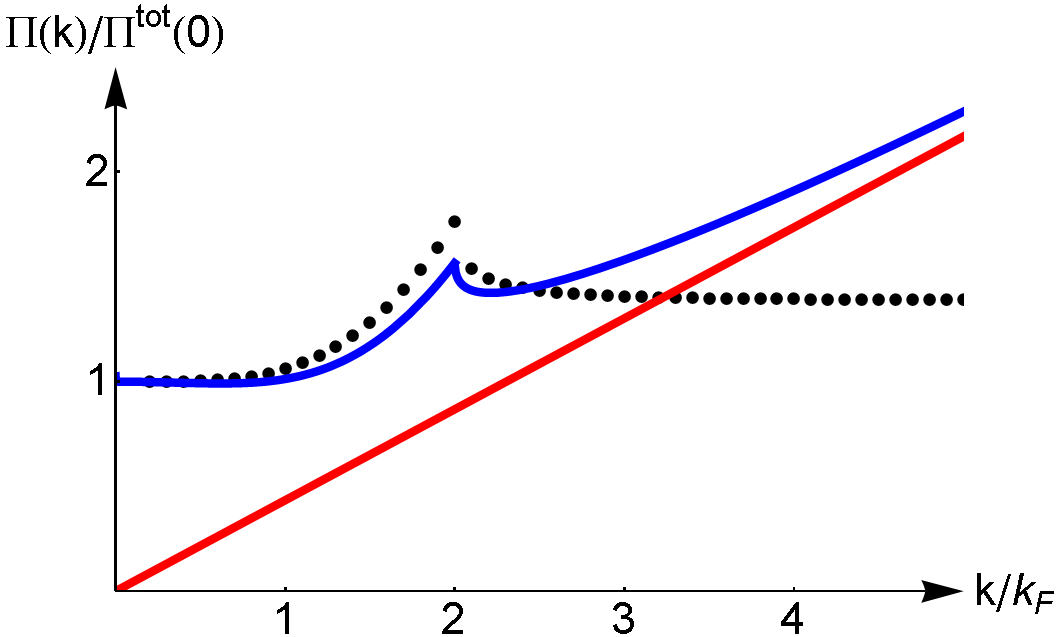}
\caption{(color online) Comparison of our results for the bilayer with the exact results.\cite{Gamayun2011} Black dotted line are the results from the two-band model. Solid linear (red) line is the linear part of the polarization and the (blue) nonlinear line is the exact result. This plot is for $k_F^2/t_\perp^2=0.1$.}
\label{fig7.e}
\end{figure}

In Fig.~\ref{fig7.e}, a direct comparison between the exact results from Gamayun\cite{Gamayun2011}(blue nonlinear line) and our approximation for a graphene bilayer is shown for $k_F^2/t_\perp^2=0.1$. As expected, our asymptotic results agree with his in the small (dotted line) and large (red linear line) momenta regimes. On the other hand, if we compare our results with Ref.~\onlinecite{Hwang12}, we do not find a good agreement. For a doping level $n \sim 1 \times 10^{12}$, similar to the one used to derive Fig.~\ref{fig7.e}, the results shown in Fig.~2 of Ref.~\onlinecite{Hwang12} disagree with both ours and Gamayun's exact result for the bilayer. Although the curves have the same qualitative shape, exhibiting a small cusp at $k=2k_F$ (their cusp peaks at a higher value compared to the exact results), the onset to the linear regime is not clearly seen in their figure. A linear fit to their large-$k$ results does not extrapolate through the origin, as it should. Although this transition could take place at larger $k$-values than those shown in their figure, the linear regime is clearly seen in the results of Gamayun, within the parameter range considered by them, especially for large electron densities ($n \sim 10^{13}$ cm$^{-2}$). This absence of a linear regime in Ref.~\onlinecite{Hwang12} occurs for bilayers, as well as for trilayers. This is an important difference, because for the trilayer we expect the linear regime to become dominant in more or less the same region as in the bilayer case. Although in our Fig.~\ref{fig7.6a} the Fermi energy is chosen to be quite high to emphasize the effect, and the transition to the linear regime shifts to larger momenta when the electron doping is reduced, there is no sign of the linear regime in the trilayer results of Ref.~\onlinecite{Hwang12}. The validation of our findings by comparison with exact results for the bilayer makes us confident that the linear behavior for the trilayer, which we have obtained numerically, is a valid and important feature. Moreover, our method should be especially valuable for evaluating the screening in systems with a higher number of layers, when exact numerical calculations become very time consuming.

The unscreened Coulomb potential $V(k)=g/k$ with interaction strength $g$ diverges as $k \to 0$. If $k_F=0$, the polarization diverges as well for $k \to 0$. The screened potential will converge to zero in this limit, $V^\textrm{tot}(k \to 0)=0$. When $k_F \neq 0$, the potential will be finite and nonzero everywhere. The screened potential has a local minimum and will converge to a renormalized version of the unscreened potential in the large-$k$ limit. Due to the linear behavior of the polarization $\Pi^\textrm{tot}(k)= -\gamma k$ at large momenta, the screened potential is just the unscreened one with a renormalized interaction strength in this regime. The new interaction strength has the form $$ \tilde{g}=\frac{g}{1+\gamma}.$$ Since $\gamma$ is positive, the interaction strength will be reduced. A sketch of the screened potential is shown in Fig.~\ref{fig7.6b}.

With the insights gained here, we conclude that the simplified model for the polarization proposed in Ref.~\onlinecite{Rvg13} to include the effect of screening in the ferromagnetic phase transition in ABC-stacked trilayer can be further refined. Indeed, instead of using the slope for the bilayer, $\gamma=0.125$, we have found that the precise value for trilayers should be $\gamma=0.18$. As a result, the critical couplings, and therefore also the critical doping levels, are reduced. The reduction of the interaction parameter $g$ becomes larger when $g$ increases. For $g=6$ the extra reduction is around $28\%$. Since the phase boundary $n_\textrm{crit}(g)$ is quite flat for $g>1$, the critical curve of the screened case in the phase diagram will not change drastically.

In sum, the results presented here allow one to determine the effect of (static) screening in $N$ layers of ABC-stacked graphene and should thus contribute to more accurate calculations of interaction phenomena in multilayer graphene. Our results for the bilayer are a good approximation of the exact expression derived by Gamayun.\cite{Gamayun2011} Since the small momentum behavior of the polarization is governed by the two-band model and the high-momentum part of the polarization is caused by the linearity of the energy bands further away from the Dirac point, our results are intuitively correct. Although a direct comparison between our results and those of Ref.~\onlinecite{Hwang12} is difficult, because in their studies the linear regime has not yet been reached, we expect from our model that their results should already start converging towards the linear asymptote for the regime of parameters considered. Since the results of this paper are valid for any ABC-stacked multilayer system, they allow for a realistic inclusion of screening phenomena in multilayer systems in a way that all the characteristics of the polarization function are included. The number of trilayer and multilayer experiments is increasing over the last few years, hence theoretical work is needed as well. Our results, which connect the universal low-momentum part and the high-momentum linear regime of the polarization, could be a useful tool to include screening in a more realistic and practical manner in the description of these fascinating materials.

\section*{Acknowledgments}
The authors acknowledge financial support from the Netherlands Organization for Scientific Research (NWO), as well as useful discussions with L. Fritz and useful correspondence with O.V. Gamayun. The authors thank V. Juri\u{c}i\'{c} for proof reading the paper.

\begin{appendices}

\section{Green's function}

In order to derive the non-interacting Green's function for ABC-stacked multilayer graphene, we can use the Feynman path integral formalism. Using the definitions in Eq.~\ref{FulLHam}, the partition function is given by
$$Z=\int \textrm{d}[\psi^\dagger] \textrm{d} [\psi] e^{-S[\psi^\dagger,\psi]/\hbar},$$ where the action is given by
\begin{align}
\nonumber S&=\int_0^{\hbar \beta} \textrm{d} \tau \int d^2\mathbf{k} \psi^\dagger(\mathbf{k})  \left( \hbar \frac{\partial}{\partial \tau} +\mathcal{H}_0(\mathbf{k}) \right) \psi(\mathbf{k}).
\end{align}
In this expression, $\tau$ denotes imaginary time. Next, define $U_\mathbf{k}$, such that $D(\mathbf{k}) \equiv  U^\dagger_\mathbf{k} \mathcal{H}_0(\mathbf{k}) U_\mathbf{k}$ is diagonal and unitary. $U_\mathbf{k}$ is called the \textit{diagonalizer} of the Hamiltonian. For the full-band model it can only be determined numerically. The diagonalizers induce a change of basis that is used later $\varphi(\mathbf{k})=U \psi(\mathbf{k})$.
\begin{align}
	\label{phiofk} \varphi^\dagger_{\mathbf{k},\sigma} &\equiv  \left( c^\dagger_{\mathbf{k},\sigma,1} , c^\dagger_{\mathbf{k},\sigma,2} , .\ ... .\  , c^\dagger_{\mathbf{k},\sigma,2N-1} , c^\dagger_{\mathbf{k},\sigma,2N} \right),
\end{align}
where $c^\dagger_{\mathbf{k},\sigma,\alpha}$ ($c_{\mathbf{k},\sigma,\alpha}$) creates (annihilates) an electron with momentum $\mathbf{k}$, spin $\sigma$ in energy band $\alpha$. The Hamiltonian in Eq.~(\ref{FulLHam}) can be rewritten as
\begin{equation}
	\mathcal{H}_0= \sum_{\mathbf{k},\sigma} \varphi^\dagger_{\mathbf{k},\sigma} D(\mathbf{k}) \varphi_{\mathbf{k},\sigma}.
	\label{nonintdiscrham}
\end{equation}

The Green's function is defined as $\langle \psi^\dagger(\mathbf{k}) \psi(\mathbf{k}) \rangle$, which is in fact equivalent to the inverse of the quadratic part of the action. After performing a Fourier transformation from imaginary time to Matsubara frequencies and neglecting spin, this results into
\begin{align}
\nonumber G(i \omega_m, \mathbf{k}) &= \left[ i \omega_m \mathbb{1}-\mathcal{H}_0(\mathbf{k}) \right]^{-1}, \\
\nonumber &= \left[ i \omega_m \mathbb{1}-U_\mathbf{k} D(\mathbf{k}) U^\dagger_\mathbf{k} \right]^{-1}, \\
\nonumber &= \left\{ U_\mathbf{k}  \left[ i \omega_m \mathbb{1}-D(\mathbf{k}) \right] U^\dagger_\mathbf{k} \right\}^{-1}, \\
\nonumber &= U_\mathbf{k}^\dagger  \left[ i \omega_m \mathbb{1}-D(\mathbf{k}) \right]^{-1} U_\mathbf{k}, \\
\nonumber &= \sum_{s=1}^{2N} \frac{1}{i \omega_m-\xi_\mathbf{k}^s} U_\mathbf{k}^\dagger  \Delta^s U_\mathbf{k},
\end{align}
where $\xi_\mathbf{k}^s$ are the diagonal entries of $D(\mathbf{k})$, which are the eigenenergies of the Hamiltonian, and $\Delta^s$ is a $2N \times 2N$ matrix with zero's everywhere, except for a $1$ in the $s^{\textrm{th}}$ entry along the diagonal. In other words, the sum over $s$ is over the different energy bands. This Green's function is a $2N \times 2N $ matrix. The components correspond with sublattice index. Hence, the propagator between two different sites $i$ and $j$ is defined as
\begin{align}
\nonumber G_{ij}(i \omega_m, \mathbf{k}) &=  \sum_{s=1}^{2N} \frac{1}{i \omega_m-\xi_\mathbf{k}^s} \left( U_\mathbf{k}^\dagger  \Delta^s U_\mathbf{k} \right)_{ij}.
\end{align}

\end{appendices}

\end{document}